\documentclass[conference]{IEEEtran}

\usepackage{cite}
\usepackage{amsmath,amssymb,amsfonts}
\usepackage{mathabx}
\usepackage{algorithmic}
\usepackage{subcaption}
\usepackage{graphicx}
\usepackage{textcomp}
\usepackage{xcolor}
\usepackage{float} 
\usepackage[font={small}]{caption}

\newcommand{\curlb}[1]{\left\{#1\right\}}
\newcommand{\squareb}[1]{\left[#1\right]}
\newcommand{\eX}[1]{e^{#1}}

\begin{document}
\title{Vector Quantized Information Bottleneck Mitigation of Adversarial Attacks on Modulation Classifiers} 

\author{
\IEEEauthorblockN{\normalsize Attanasia Garuso, Silvija Kokalj-Filipovic}
\IEEEauthorblockA{\small Rowan University\\
\small\em \{garuso38, kokaljfilipovic\}@rowan.edu}}

\maketitle
\begin{abstract}
Radio-frequency (RF) signal processing based on machine learning, including Automatic Modulation Classification (AMC), plays a critical role in Next Generation (NextG) intelligent communications but remains highly vulnerable to adversarial attacks. Prior work explored Variational Information Bottleneck (VIB) as a means to improve robustness, but only in continuous spaces and never for RF classifiers. We investigate the impact of discrete-space information bottlenecks on the adversarial robustness of AMC models. Classification accuracy of attacked modulated RF waveforms is evaluated in two frameworks designed to leverage VIB for adversarial robustness: (i) a two-stage framework where VIB is applied in the preprocessing step using a Vector Quantized Variational Autoencoder (VQVAE) to suppress adversarial perturbations in the process of input signal reconstruction and (ii) a Vector Quantized Classifier (VQC) that leverages VIB inside the classifier proper. The  architectures of VQVAE and VQC are identical up to the bottleneck where the VQVAE adds a decoder while VQC utilizes a classification head. We train both VQVAE and VQC  with loss functions that we derive from the VIB information-theoretic model. Both frameworks incorporate vector quantization and KL regularization to enforce discrete bottlenecks that filter out adversarial noise. Evaluations under white-box attacks demonstrate that discrete bottlenecks significantly enhance robustness: VQVAE outperforms its continuous VAE (Variational Autoencoder) counterpart, while VQC achieves strong robustness without requiring signal reconstruction.  
\end{abstract}
\vspace{-4mm}
\section{Introduction}
Adversarial attacks on machine learning models pose a significant challenge in RF signal processing, a domain that is critical for NextG intelligent cellular communication and spectrum sensing. Baseband signal processing can be contained within individual commercial devices (sensors, IoT devices, smartphones) and experimental USRP platforms, or distributed across many Open-RAN components \cite{d2022dapps} and collaborative spectrum sensors. The Open-RAN AI algorithms will require substantial data transfer between its components, even at the level of baseband signals \cite{d2022dapps}. This calls for careful consideration of data vulnerabilities given the multi-vendor nature of Open-RAN and  the sheer amount of digitally-modulated baseband data for the training of NextG AI systems.  We investigate white-box attacks only (i.e. those based on a known model), considering the trend of open-source ML models available through an increasing number
of platforms and hubs. 

Quite a lot of research has been conducted on the topic of RF adversarial attacks \cite{Adexdetect,AEagainst, sadeghi2018adversarial, realizationRFadvers}. We will here demonstrate how training with a Variational Information Bottleneck (VIB) objective \cite{alemi2019deepvariationalinformationbottleneck} makes an automatic modulation classification (AMC) model \cite{OriginalModRec, Access2021AMC, ZHANG2022AMC}, significantly more robust to adversarial examples. In particular, we direct our attention to the VIB implemented in discrete space, e.g. how a VQVAE can be applied to mitigate the attack on RF samples, received or stored by the victim. Although autoencoders have been previously used to mitigate RF adversarial examples (AdExs) \cite{AEagainst},  autoencoders trained with a Discrete-Space VIB (DSVIB) objective have not been studied as mitigation mechanisms.  VQVAE mitigation is implemented in a 2-step framework: (1) the attacked signal is passed through the VQVAE to suppress adversarial perturbations; (2)  it is the output of the VQVAE that is fed to the Classifier Under Test (CUT) instead of the attacked signal. The second DSVIB-based mitigation framework relies on using a VIB-trained classifier: Vector Quantized Classifier (VQC) leverages DSVIB inside the classifier proper. VQC shares the encoder and the vector quantizer with the VQVAE from the $1$st framework, but the VQVAE decoder is replaced by a classification head. Both VQVAE and VQC are trained using loss functions derived from the VIB information-theoretic model. For fair comparison, the CUT used in the $1$st framework has the same architecture as VQC, with the exception of vector quantizer, and it is trained independently.

While acknowledging the relevance and complexity of an over-the-air attack \cite{realizationRFadvers}, which demands precoding for multipath equalization, this paper only presents attacks on the clean data. The clean data comes from high-SNR signals, obtained after channel estimation and equalization. This illustrates the case where baseband data is moved between the Open-RAN components for signal processing or AI training, or stored at the edge, and is thereby exposed to an attack off-the-air. Beyond AMC, new data-hungry AI-based channel models \cite{MIMOChannelEstimScore,ScoreDiffChannel,yu2024ai} would be negatively affected by such data poisoning, making the proposed attack suppression an important contribution.

Our main contributions are as follows: 1. Derivation of the VIB-based training objectives with the bottleneck in the discrete space applied to two distinct  models (VQVAE and VQC) 2. Implementation of VQVAE and VQC processing frameworks of the RF baseband data, which enables fair comparison of their effects on the adversarial robustness of CUT 3. Quantification of CUT robustness against two attack types based on classification accuracy with and without VIB-based mitigation  4. Discrete latent space analysis of VQVAE and VQC using set distances to quantify attack-induced shifts in discrete mappings of the signals with modulation-specific insights (Fig.~\ref{fig:set_dist}). 
The paper is organized as follows: Section~\ref{sec:VIBintro} introduces the frameworks and derives the VIB-based objectives. Section~\ref{sec:AMCattacks} describes the RF data, classification model and adversarial attacks. Section \ref{sec:method} describes how the methodology of defending against RF adversarial attacks is implemented. Section \ref{sec:evalmethod}  outlines evaluation methods, and discusses their results. Finally, we conclude with insights and directions for future research.
\vspace{-1.5mm}
\section{Problem Definition} \label{sec:VIBintro} 
It is well known that autoencoders mitigate adversarial attacks on classifiers by leveraging the latent input representation in its bottleneck \cite{dimred, AEagainst}. Previous work has been conducted which analyzes this mitigation from an information-theoretic perspective using VIB \cite{alemi2019deepvariationalinformationbottleneck, IBrobust, AdvRobICMLCN1}. In this paper, we investigate whether the implementation of VIB via a discrete-space  bottleneck (DSVIB) further enhances robustness compared to standard autoencoders and continuous VIB implementations. We also analyze the impact of embedding this bottleneck directly within the classifier. Specifically, we study how a discrete bottleneck affects robustness across different attack types, and how this translates to the protection of baseband RF samples from six distinct waveform classes.
\vspace{-1.5mm}
\subsection{Classifier Under Test Model}
\begin{figure}[h]
\centering
\includegraphics[width=0.48\textwidth]{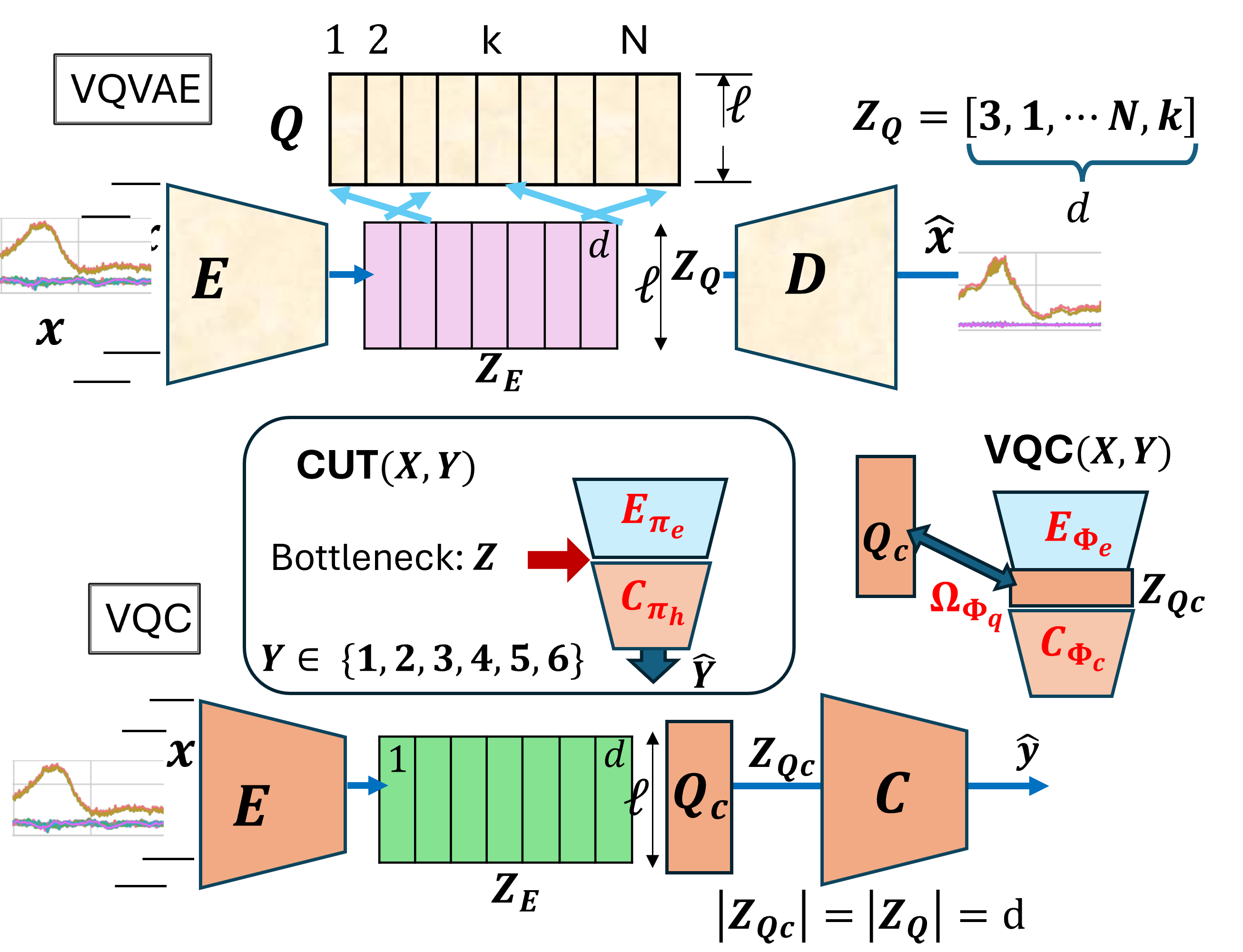}
\caption{Architecture of 2 DSVIB-based models: VQVAE and VQC, and the matching AMC CUT. While the codebooks have the same shape, they have different parameters, hence $Q$ and $Q_c.$
}\vspace{-3mm}
\label{fig:vqmodels}
\end{figure}

The CUT serves as the baseline architecture for evaluating the effects of vector quantization (VQ) on adversarial robustness. We will compare the attacks on CUT against the attacks on similar classifiers hardened by DSVIB. CUT consists of two main components: a feature extractor (FE) and a classification head (CH). The FE $E_{\pi_e}$ maps the input waveform $x$ into a latent representation $Z$, while the CH $C_{\pi_h}$ maps $Z$ to the target label $\hat{y}$ (Fig.~\ref{fig:vqmodels}).
CUT represents the standard end-to-end classifier without any quantization or reconstruction modules, and with no other mitigation implemented. It provides a reference point for assessing how the introduction of a DSVIB via VQVAE or VQC affects AMC robustness.

Both VQVAE and VQC are derived directly from the CUT architecture. Specifically, as shown in Fig.~\ref{fig:vqmodels}:
\begin{itemize}
    \item VQVAE reuses the CUT’s feature extractor as its own encoder, and adds vector quantization (VQ) and a decoder to form the reconstruction model used as a pre-processing step.
    \item VQC also repurposes the CUT’s FE as its own encoder $E_{\phi_e}$ and introduces VQ before the CH $C_{\phi_c}$, forming a discrete-latent classifier.
\end{itemize}
Therefore, the CUT is both the baseline classifier and the structural foundation for the proposed VIB-based models.
\vspace{-1mm}
\subsection{VIB-based Models}
While the VIB principle can, in theory, be applied to any deep learning model, we focus on two architectures: the VQVAE, which reconstructs signals $X$ through a discrete latent space $Z_q$, and the VQC, which performs classification directly on the quantized latent $Z_{q_c}$. Our analysis seeks to understand how these models leverage the DSVIB to mitigate adversarial perturbations on RF data classified by the CUT.

Before introducing each model individually in Subsections~\ref{subsec:vqvaeimpl} and \ref{subsec:vqcimpl}, it is important to describe how these architectures are employed within the adversarial mitigation framework and to clarify their conceptual distinctions. In the case of VQVAE, the model is used as a pre-processing step prior to classification: before an attacked waveform is fed into CUT, it is first passed through the VQVAE. The model has an encoder and a VQ, followed by a decoder.  Encoder maps $X$ into continuous latent representation $Z_e,$ which is  quantized into a discrete latent $Z_q$ using a trained codebook $Q$. As VQVAE reconstructs $X$ by passing $Z_q$ through its decoder, it is $Z_q$ that filters out adversarial perturbations as insignificant. The reconstructed waveform $\hat{X}$, which retained the salient information only, is then passed to the classifier for the final task.

In contrast, the VQC does not have a separate pre-processing step since the quantization is built into the classification model which is being attacked. 
Our VQC model consists of encoder, vector quantizer and classification head, where the former two have the same architecture as the respective elements of VQVAE to generate the identically shaped bottleneck.  We therefore rely on this built-in discrete space mapping to automatically mitigate the effects of an attack if one occurs. 

While both approaches rely on vector quantization, the way the bottleneck is trained is different, i.e., task-aligned. In the VQVAE, the bottleneck is trained on the task of reconstruction - it learns to make the output identical to the input, capturing in the bottleneck only the essential information needed for reconstruction. In the VQC, the bottleneck is trained on the task of classification. The VQ is optimized not to reproduce the input, but to improve  classification performance while maintaining its robustness. In conclusion, two distinctive differences between VQVAE and VQC lay, first, in how the models are used within the mitigation pipeline, and second, in the task-aligned training.
\subsection{Theoretical Model}
To formalize DSVIB, we start with Shannon’s Rate-Distortion (R-D) theory~\cite{ShannonCoding} which seeks to optimize  lossy compression of a random variable (r.v.) $X,$ for a given  distortion constraint. Formally, for a distortion measure $\mathcal{D}$ and a distortion constraint $D_c$, the minimal bit-rate is characterized by the \textit{rate-distortion function}:
\begin{equation}
\underset{P_{Y|X}}{\mathrm{min}}\; I(X;Y)\; \text{s.t.}\; \mathcal{D}(X,Y) \leq D_c\label{eq:rdfunction}
\end{equation}
where $I(X;Y)$ is mutual information, defined as
\begin{equation}
\mathrm{I}(X;Y)=\int_{}\int_{} p(x,y) \log \left(\frac{p(x,y)}{p(x) p(y)}\right) dxdy\label{eq:mutualinf}
\end{equation}
 The R-D objective translates to Information Bottleneck (IB) \cite{TishbyIB} that maps $X$ to a hidden representation $Z$, which is minimally informative of $X,$ hence of low bit-rate, while maximally informative of the target r.v. $Y,$ by maximizing the IB objective function
\begin{equation}
J_{IB}=I(Z;Y) - \beta I(Z;X)\label{eq:vib},
\end{equation}
where $\beta$ is the Lagrange multiplier that multiplies the so called complexity term, wrapping in the bit-rate minimization objective from \label{eq:rdfunction}~\cite{singhLearnedComp}. A large $\beta$ enforces more compression. 
For example, given a classification model of $X$, and ground-truth labels $Y$ from a joint distribution $P_{X,Y}$, the IB objective is to maximize $J_{IB}$ over $P_{X,Y,Z}$ which form a Markov chain derived from the position of the latent representation $Z$ within the classification model (e.g., at the output of $E$ in CUT, Fig.~\ref{fig:vqmodels}).
Essentially, IB  forces $Z$ to be a minimal sufficient statistic of $X$ for predicting $Y.$ The main drawback of the IB principle is that computing mutual information is, in general, computationally challenging \cite{EstimI}. VIB is an extension of IB tied to deep-learning models. VIB uses variational inference to approximate the intractable mutual information terms in the IB objective function. 
%
Note that computing the marginal distribution of $Z$ in $I(X;Z)$ (see Eq.~\ref{eq:mutualinf}), 
\begin{equation}	
p(z) = \int{p(z \mid x)\, p(x)\, dx},\label{eq:pz}
\end{equation}
is particularly difficult. The Variational Inference approach replaces the posterior $p(z\mid x)$, the updated belief about the latent $z$ after observing data $x,$ with an approximation function $q_{\theta} (z\mid x)$  parameterized by an encoder model $E(x,\theta)$ with parameters $\theta$.  $q_{\theta} (z\mid x)$    is trained to resemble $p(z\mid x)$  through a loss function which contains the Kullback-Leibler (KL) divergence  $KL(p(z\mid x)||q_{\theta} (z\mid x)$.

 As we are interested in the discrete latent space $Z_q$ obtained by VQ of the information bottleneck r.v.  $Z,$  we now derive the VIB loss functions for the training of VQVAE and VQC represented in Fig.~\ref{fig:vqmodels}.
\subsection*{Task-aligned Loss Functions}
For both models, the joint probability of the r.v.-s in Eqn.\ref{eq:vib}  leverages their Markov chain
relationship:
\begin{align}
p(X, Y, Z) 
\nonumber &= p(Z \mid X, Y)\, p(Y \mid X)\, p(X) \\
&= p(Z \mid X)\, p(Y \mid X)\, p(X),  \label{eq:jnt}
\end{align} 
where $X$ is the input signal, $Z$ is the bottleneck representation, and $Y$ is the task-aligned output.
\subsubsection{\bf VQ-Classifier}
Let us denote the components of VQC as $E_{\phi_e}$, $Q_{\phi_q}$ and $C_{\phi_c},$ where $\phi_{\cdot}$ denotes respective trainable parameters of each component.
For the $I(Z; Y )$ factor in Eqn.~\ref{eq:vib}, replacing  $p(y\mid z)$ by its variational approximation $q(y\mid z)$, which represents probabilistic decision y of the classifier module $C_{\phi_c} (z_q )$, we obtain the inequality
\begin{align}
I(Z; Y) 
&\geq \int p(y, z) \log \frac{q(y|z)}{p(y)} \, dy \, dz \nonumber \\
&= \int p(y, z) \log q(y|z) \, dy dz - \int p(y, z) \log p(y) \, dy dz  \nonumber \\
&= \int p(y, z) \log q(y\mid z)\, dy \, dz + H(Y). \label{eq:11}
\end{align}
For simplicity, we use $z$ instead of $z_q,$ the quantized input to $C_{\phi_c},$ which we will factor in later. As the entropy of the labels, $H(Y)$, is independent of the VIB optimization, it can be ignored.  Focusing on the first term in Equation~\eqref{eq:11}, we can rewrite
\[
p(y, z) = \int p(x, y, z)\, dx = \int p(x)\, p(y \mid x)\, p(z \mid x)\, dx
\]
(leveraging the Markov assumption).  
This gives us a new bound:
\begin{align}
I(Z; Y) \geq \int p(x)\, p(y \mid x)\, p(z \mid x) \log q(y\mid z)\, dx\, dy\, dz.\label{eq:12}
\end{align}
Notice that integration over $x$ is the training process itself, while $p(y\mid x)$  in Eqn.~\ref{eq:12} represents the ground truth label $y$ ($y_{GT}$) of $x.$ Overall, Eqn.~\ref{eq:12} represents the training averaging of the cross entropy (CE) loss of the VQC classifier: $L^{ce}(C_{\phi_c}(z),y_{GT})=p(y_{GT} \mid x)\times  \log q(y\mid z)$. 

In the second part, we consider the term $I(Z;X)$ of Eqn.~\ref{eq:vib}:
\begin{align}
I(Z; X) 
&= \int p(x, z) \log \frac{p(z|x)}{p(z)} \, dx \, dz \nonumber \\
&= \int p(x, z) \log p(z|x)\, dx \, dz - \int p(z) \log p(z)\, dz. \label{eq:13}
\end{align}
\vspace{-1.0mm}
To overcome difficulties in computing $p(z)$ in Eqn.~\ref{eq:pz}, we replace it with $r(z)$, a variational approximation of this marginal, implemented by the encoder $E$ followed by the quantizer $Q.$ Note that $z$ in Eqn.~\ref{eq:13} stands for the quantized  $z_q$. 
Since $\mathrm{KL}[p(Z) \| r(Z)] \geq 0$, it follows that
\[
\int p(z) \log p(z)\, dz \leq \int p(z) \log r(z)\, dz,
\]
which gives the following upper bound:
\begin{align}
I(Z; X) \leq \int p(x)\, p(z \mid x) \log \frac{p(z|x)}{r(z)} \, dx \, dz. \label{eq:14}
\end{align}
The above integration over $x$ is performed by leveraging stochastic gradient descent-based training of VQC on the dataset $X$ of length $T.$ The $I(Z;X)$ in the VIB Eqn.~\ref{eq:vib} is to be minimized, which is done with the KL loss function $\mathrm{KL}\big( p(z \mid x) \,\|\, r(z) \big)$, 
giving the implementation of Eqn.~\ref{eq:vib}:
\begin{align*}
&\mathcal{J}_{vib} = \tfrac{1}{T} \sum_{n=1}^T \mathbb{E}_{sg} \Big[ \log q(y_n \mid z) 
- \beta\mathrm{KL}\big( p(z \mid x_n) \| r(z) \big) \Big]= \\
&\tfrac{1}{T} \sum_{n=1}^T \mathbb{E}_{sg} \Big[ -L^{ce}(C_{\phi_c}(z) \mid x_n, y_n) 
- \beta\, \mathrm{KL}\big( p(z \mid x_n) \| r(z) \big)  \Big],
\end{align*}
for the continuous space bottleneck (as in \cite{alemi2019deepvariationalinformationbottleneck}).
For the discrete case (our setup), it becomes
\begin{align}
\nonumber &\mathcal{J}^{vqc}_{vib} = \tfrac{1}{T} \\
&\sum_{n=1}^T \mathbb{E}_{sg} \Big[ - L^{ce}(C(z_q) \mid x_n, y_n) 
- \beta\, \mathrm{KL}\big( p(z_q \mid x_n) \,\|\, r(z_q) \big)\Big].   \label{eq:jc}
\end{align}
\vspace{-0.5mm}
Here $\mathbb{E}_{sg}$ is the stochastic averaging by training. Hence, the loss function for training VQC is $\mathcal{L}^{vqc}_{vib} =  L^{ce}(C(z_q) \mid x_n, y_n) 
+ \beta\, \mathrm{KL}\big( p(z_q \mid x_n) \,\|\, r(z_q) \big)$.  
Note above that the distribution $p(z_q |x_n)$ is {\bf discrete}, i.e. we perform stochastic vector quantization of each slice $j$ of the encoder $E$’s output $z$ by sampling  
\[
p(z^j_q=k|x) = \eX{-\big\| E(x, \phi_e)[j] - e_k \big\|^2},
\]
$E_{\phi_e} (x)$ is output of the encoder module, while  $e_k$ is the $k$th codeword (CW) of the trainable codebook $Q_{\phi_q}$ (Fig.~\ref{fig:vqmodels}). Each of the $d$ slices of $z$ is a vector of the same length $\ell$ as the codewords of $Q_{\phi_q}.$  Because both $E$ and $Q$ are trainable, the KL loss in Eqn.~\ref{eq:jc} effectively creates two more loss components (due to partial gradients with respect to both modules): the commitment loss $L_c  = \sum_j{\big\| E(x, \phi_e)[j] - F(e_k) \big\|^2}  $  and the quantization loss $L_Q  = \sum_j{|\big\| F(E(x, \phi_e))[j] - e_k \big\|^2}.$ The $F(\cdot)$ is an operator that denotes a frozen neural net (with its current parameters fixed). The training over $L_c$ and $L_Q$ is done via Gumbel-Softmax variational approximation of the stochastic mapping between $z$ and $z_q,$  $p(z^j_q=k|x).$  Hence, the final loss function implementation is 
\vspace{-1.5mm}
\begin{align}
\nonumber &\mathcal{L}^{vqc}_{vib} =  L^{ce}(C(z_q) \mid x_n, y_n) 
+ \\
&\beta\,\curlb{\mathrm{KL}\big( p(z_q \mid x_n) \,\|\, r(z_q) \big)+L_c+L_Q}. \label{eq:lc}
\end{align}
Note that the prior in the KL component is the uniform distribution over the CW indices $k\in \curlb{1,\cdots, N}.$
\subsubsection{\bf VQ-VAE}
For VQ-VAE, the distribution $p(X, Y, Z)$ in Eqn.~\ref{eq:jnt} of the VIB model trained on the reconstruction task is customized with 
$Y=\hat{X},$  the output aiming to reconstruct $X$ given $Z.$ 
Hence, Eqns.~\ref{eq:12} and \ref{eq:14} can be applied to the case of VQVAE by interpreting $Y$ as $\hat{X},$ and $q(y\mid z)$ as the decoder neural net $D_{\theta_d}$. Note that the integral of $p(z \mid x) \log q(y\mid z)$ in Eqn.~\ref{eq:12} is the minimum mean square error loss (MMSE) between $x$ and $y=\hat{x},$ and $p(y \mid x)$ is an identity. Hence, the loss for the VQVAE becomes:
 \begin{align}
\nonumber &\mathcal{L}^{vqv}_{vib} =  MMSE(D_{\theta_d}(z_q), x_n) 
+ \\
&\beta\,\curlb{\mathrm{KL}\big( p(z_q \mid x_n) \,\|\, r(z_q) \big)+L_c+L_Q}. \label{eq:ld}
\end{align}
\vspace{-1.0mm}
The loss $\mathcal{L}^{vqv}_{vib}$ is used to jointly train all three components of the VQVAE model, $E_{\theta_e}$, $Q_{\theta_q}$ and $D_{\theta_d}$. 

\section{AMC Model and Adversarial Attacks}\label{sec:AMCattacks}
The CUT model that was introduced in section~\ref{sec:VIBintro} is trained to perform Automatic Modulation Classification (AMC) on the dataset described below. This section introduces the AMC model in terms of both the CUT architecture and the data used for the training and evaluation. 
\vspace{-1.5mm}
\subsection{Dataset}\label{subsec:dataset}
The dataset used for this study consists of datapoints which represent sampled RF signals, of the types common in modern wireless communications. Each datapoint represents a sampled complex-valued modulated RF waveform arising from a specifically modulated sequence of random information bits, converted to a baseband RF signal. A datapoint $x$ can be represented as $x=\squareb{Re_i+j Im_i}, i = 1\cdots p$ with $j=\sqrt{-1}$.
First, a modulated signal
$u$ is obtained as $u = M_s(b)$, where $s \in \mathcal{S}$ is the employed
digital modulation scheme. For any $s$, $M_s = \curlb{0,1}^{m(s)} \rightarrow \mathcal{C}^v$ describes the modulation function which encodes random bit sequence $b$ of length $m$ into a sequence $u$ of complex valued numbers of length $v.$ We here extended the notion of modulation, as we included the OFDM waveform:
\vspace{-1.5mm}
\[
\mathcal{S} = \{ \text{4ASK}, \text{8PAM}, \text{16PSK}, \text{32QAM-X}, \text{2FSK}, \text{OFDM256} \}.
\]
We create datapoints as sub-sequences  $x$ of $u \in \mathcal{C}^v,$ of length $p = 1024,$ and implement  the training dataset $X_{train}$ by using an open-source library {\em torchsig} featured in \cite{torchsig}. $X_{train}$ contains RF samples of high SNR, for the sake  of simplicity and  to match our use case. The torchsig library function {\em ComplexTo2D} is used to transform vectors of complex-valued numbers into  2-channel datapoints, traditionally referred to as the {\em I} and {\em Q} component. Each channel is comprised of $p$ real numbers, previously normalized. Channel 1 contains real components $I\in \mathbb{R}^p$ and channel 2 contains the imaginary ones $Q\in \mathbb{R}^p.$  
Modulation diversity in $\mathcal{S}$ spans both amplitude-based (PAM, ASK) and phase-based (PSK, QAM, OFDM) signals, providing a comprehensive setting for testing adversarial robustness across distinct waveform characteristics. 
For more detail, please see \cite{AttaFCN}.
\subsection{Classifier Architecture}
The CUT architecture comprises an encoder and a classification head. The encoder maps the two-channel waveform input $x$ into a latent representation $Z_e$ using three batch-normalized 1D-convolutional layers and two residual blocks, and the classification head maps $Z_e$ into the predicted label $\hat{y}$ out of 6 classes:
\vspace{-1.5mm}
\[
Z_e = E(x, \pi_e), \quad \hat{y} = C(Z_e, \pi_h).
\]
\vspace{-1.0mm}
This model provides the foundation for both the VQC and VQVAE-based architectures. 
\vspace{-1.5mm}
\subsection{Adversarial Attack Methods} \label{subsec:attacks}
We evaluate robustness using two adversarial attacks, FGSM and PGD. All attacks directly perturb the input data $x$ (composed of $I$ and $Q$ channels) and produce adversarial examples $x_a,$ which we use to test classifiers.
For the Fast Gradient Sign Method (FGSM), a classifier with weights $\pi$ (or $\phi$ in VQC) is attacked by modifying the input $x$ imperceptibly while maximizing the loss. Perturbations of low strength $\epsilon\leq 0.2$ are added in the direction of the loss gradient:
\vspace{-1.5mm}
\[
x_a = x + \epsilon\,\mathrm{sign}\!\big(\nabla_x J(\pi,x,y)\big),
\]
where $J$ is the cross-entropy loss. For our two-channel inputs ($I,Q$) with 1024 samples each, FGSM is applied independently on each channel:
\[
I_a = I + \epsilon\,\mathrm{sign}(\nabla_I J),\qquad Q_a = Q + \epsilon\,\mathrm{sign}(\nabla_Q J).
\]
For PGD, the same gradient-direction principle is applied iteratively and the perturbed sample is projected back into the $\epsilon$-ball around $x$ to ensure the adversarial example remains within the allowed budget. Because PGD performs multiple updates and projections, it is typically more effective than the single-step FGSM:
\[
x^{(t+1)}=\Pi_{B_\epsilon(x)}\!\Big(x^{(t)}+\alpha\,\mathrm{sign}\!\big(\nabla_x J(\pi,x^{(t)},y)\big)\Big),
\]
where $\Pi_{B_\epsilon(x)}$ denotes projection to the $\ell_\infty$ ball of radius $\epsilon$ and $\alpha$ is the step size.

Each attack is evaluated across a range of perturbation strengths $\epsilon,$ from insignificant (.05) to barely perceptable (.2). 
\vspace{-3mm}
\section{Methodology} \label{sec:method} 
We describe the implementation of the VIB-optimized VQ, which will be used to evaluate adversarial robustness in the classification of RF waveforms, under the two approaches that we introduced: the reconstruction-aligned VIB in VQVAE and the classification-aligned VIB in VQC. 
\subsection{VQVAE-Based Implementation} \label{subsec:vqvaeimpl}
VQVAE \cite{vqvae2017neural} extends the standard Variational Autoencoder \cite{KingmaVAE} by introducing vector quantization of the latent bottleneck. The encoder maps an input waveform $x$ to a continuous latent $Z_e$, which is then discretized into $Z_q$ using a learned codebook $Q$ of size $N$. 
Each of the $d$ slices of $Z_e$ is quantized by a stochastic vector quantizer $\Omega$ to the nearest CW in $Q$, producing a compressed discrete representation of the input, as described in Fig.~\ref{fig:vqmodels}. The indices of the selected codewords form $Z_q,$ an integer array of length $d.$ The decoder reconstructs the waveform $\hat{x}$ from $Z_q$: first, it  creates $\hat{Z}_e$ by replacing every slice of $Z_e$ with the CW indexed by the respective element of $Z_q$; then $\hat{Z}_e$ is passed through the decoder built as the traditional transpose of the convolutional layers in the encoder. The exact architecture is omitted here due to space limitations, but it could be found in \cite{AttaFCN}. Practically, bypassing the quantizer, i.e. $\hat{Z}_e = Z_e,$  would make this model an autoencoder. It is the VQ and the VIB-based loss which make it a VQVAE. The training maximizes the objective to pass only the salient features through the bottleneck, hence leaving the adversarial perturbations out. The amount of salient information passed depends on the compression rate (in bits) $R_c=\frac{2p\times B}{d\log N},$ where $B=32$ for a single-precision FP representation of real numbers. Given a CUT, both $d$ and $N$ depend on the codebook design $Q(\ell, N)$, since  $d=K/\ell,\ K=prod(dim(Z_e))=512\times64.$   For the results in this paper, 
$N=128$ codewords of length $\ell=512$ were used, i.e, $Z_q$ is a sequence of 
$d=(K/\ell=64$ discrete indices between $1$ and $128$. Hence, $R_c\approx 146,$ which is a great compression, given the constraint of task alignment. Note that compressed representations can be exchanged between Open-RAN processing blocks to save bandwidth, storage and time. We tested all combinations of $(\ell,N)$ from $(64,64),$ in logarithmic scale steps, up to $(512,256).$ 
In general, $N\geq128$ produced better accuracy and mitigation effects, while the impact of $\ell$ depends on $N$ in a manner that requires further research (more points). 
Formally, the trained VQVAE model is expressed as
\begin{equation}
Z_e = E(x, \theta_e), \quad 
Z_q = \Omega(Z_e, Q, \theta_q), \quad 
\hat{x} = D(Z_q, \theta_d),
\label{eq:vqvaefwd}
\end{equation}
where $\theta_e$, $\theta_q$, and $\theta_d$ are the trained parameters of the 
encoder, codebook, and decoder, respectively.
Putting the Eqn.~\ref{eq:ld} in the context of the VQVAE architecture, we observe that the training objective combines several terms:  
\begin{itemize}
    \item \textbf{Reconstruction Loss} 
    $\mathcal{L}_{rec}=MMSE(\hat{x}_n, x_n)$ enforces fidelity between $x$ and $\hat{x}=D_{\theta_d}(z_q)$.  
    \item \textbf{Commitment Loss} $\mathcal{L}_{c}$ ensures the encoder output $Z_e=E_{\theta_e}(x)$ remains close to the chosen CW, preventing 
    the encoding of irrelevant perturbations.  
    \item \textbf{Quantization Loss} $\mathcal{L}_{q}$ ensures that codewords are trained to map  the continuous latent space with the lowest average quantization error.  
    \item \textbf{KL Regularization} penalizes how much the learned distribution over CWs differs from the prior, regularizing non-salient information flow.  
\end{itemize}
The total training objective 
$\mathcal{L}_{total} = \mathcal{L}_{rec}  + 
\beta \big(\mathcal{L}_{c} + \mathcal{L}_{q} + KL \big)$
is used to train VQVAE for 200 epochs with $\beta = 0.01$, which gave the best classification and reconstruction results.
\vspace{-2mm}
\subsection{VQC-Based Implementation} \label{subsec:vqcimpl}
The VQC practically repurposed the VQVAE architecture for direct classification of RF waveforms. It sends the quantized latent representation into the classification head identical to that of CUT. This allows VQC to serve as an end-to-end robust classifier, leveraging the discrete information bottleneck via a built-in VQ and the VIB-based training loss function. 
Formally, the VQC model is expressed as
\begin{equation}
Z_e = E(x, \phi_e), \quad 
Z_q = \Omega(Z_e, Q, \phi_q), \quad 
\hat{y} = C(Z_q, \phi_c),
\label{eq:vqc_fwd}
\vspace{-1.5mm}
\end{equation}
where $\phi_e$, $\phi_q$, and $\phi_c$ are the trained parameters of the encoder, codebook, and classifier, respectively. 
The training objective from Eqn.~\ref{eq:lc} can be viewed as analogous to the VQVAE objective, except that it replaces the reconstruction loss with the classification cross-entropy loss $\mathcal{L}_{cls}$:
$\mathcal{L}_{total} = \mathcal{L}_{cls} + \mathcal{L}_{q} +
\beta \big(\mathcal{L}_{commit} +  \mathcal{L}_{q} + KL \big)$.
Via this loss, the VQC implements a discrete space VIB that will suppress adversarial perturbations while preserving the AMC task-relevant information. 
\vspace{-1mm}
\section{Evaluation Methods and Results}\label{sec:evalmethod} 
To assess the attack mitigation by the proposed discrete VIB models, we conducted experiments on RF baseband samples from six modulation classes, described in subsection~\ref{subsec:dataset}.
Both VQVAE and VQC  were evaluated on multiple white-box attacks, as described in subsection~\ref{subsec:attacks}.
Classification robustness was measured using the following:
\begin{itemize}
    \item \textbf{Classification Accuracy:} The proportion of correctly classified adversarial and clean samples.
    \item \textbf{Set Distance} $d_S$: Relative number of CW index values that appeared in $Z_q$ due to attack, which did not exist before - to capture global shifts in the latent mapping induced by attacks.  $d_S$ is the average size of the set-difference between the CW indices in $Z_q(x)$ and those in $Z_q(x_a),$ normalized by the size of set-difference between the indices in 2 consecutive drawings of $\Omega(z|x)$: note that $d_S(z_q\sim\Omega(z|x), z^{\prime \prime}_q \sim\Omega(z|x))$ is non-zero, even without the attack, due to stochastic quantization.
\end{itemize}
\begin{figure}[h]
\vspace{-3mm}
\centering
\hspace{-1mm}\includegraphics[width=0.5\textwidth]{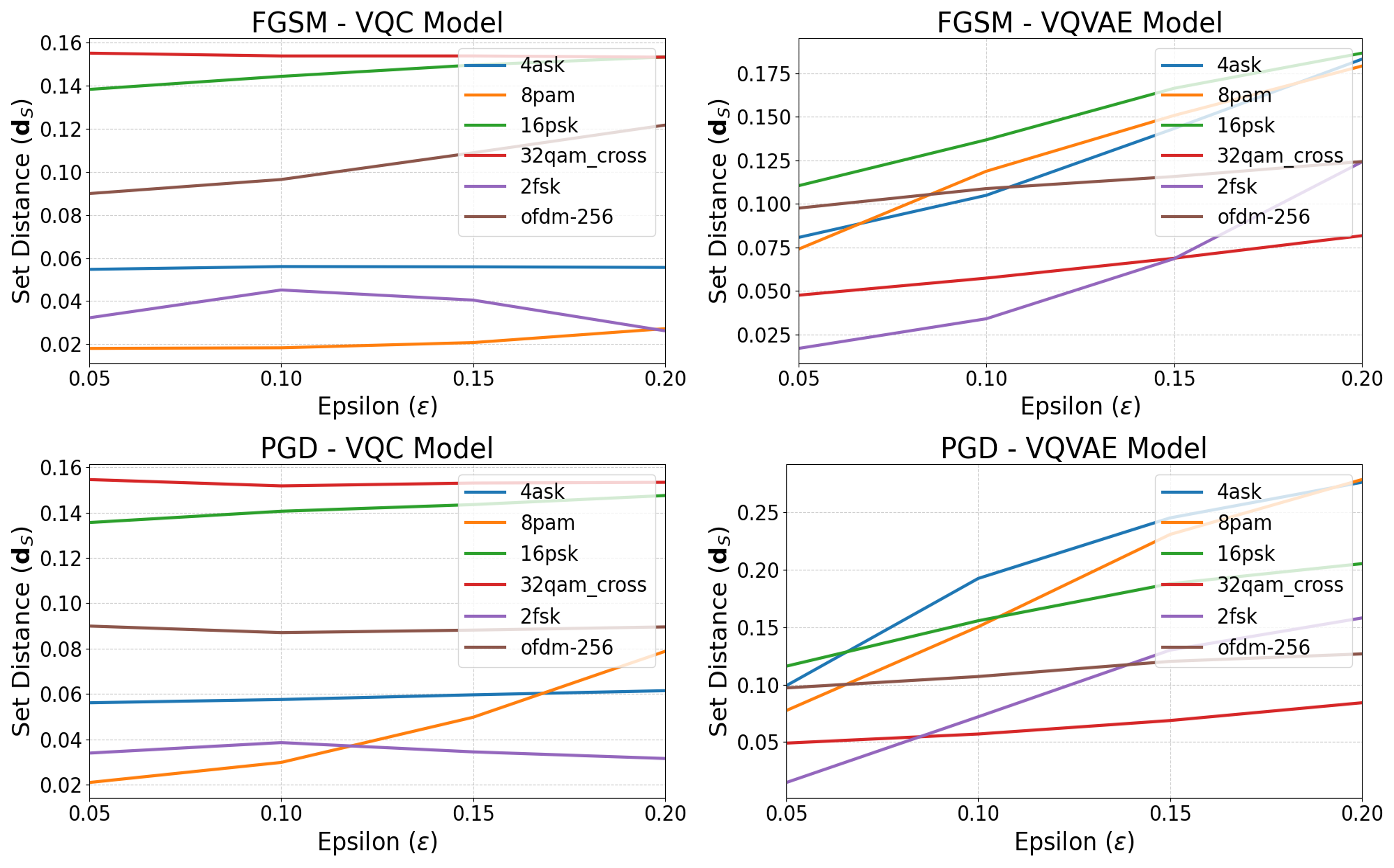}
\vspace{-3mm}
\caption{Set Distance ($\mathbf{d_S}$): Comparing sets of CW indices in $Z_q$ at VQVAE (VQC), with and without (FGSM  or PGD) attack, across modulation classes. 
}\vspace{-3mm}
\label{fig:set_dist}
\end{figure}
Fig.~\ref{fig:set_dist} shows that VQC experiences less distribution shifts in the $Z_q$ domain than VQVAE, although the stochastic nature of $\Omega(z|x)$ may require more samples for a strong conclusion (we used 1K per class). Same shifts are visible from the Hamming distances, omitted here due to space limitations (see \cite{AttaFCN} for a definition). However, VQVAE seems to mitigate the attacks (Fig.~\ref{fig:vqvae_vae}) slightly better than VQC (Fig.~\ref{fig:vqc_cut}), despite VQC being aligned with the task of classification. Most importantly, both discrete models outperform their continuous-space counterparts across all tested $\epsilon$ values. 
\vspace{-1mm}
\subsection{VQC Performance}
\begin{figure}[h]
\vspace{-3mm}
\centering
\hspace{-1mm}\includegraphics[width=0.46\textwidth]{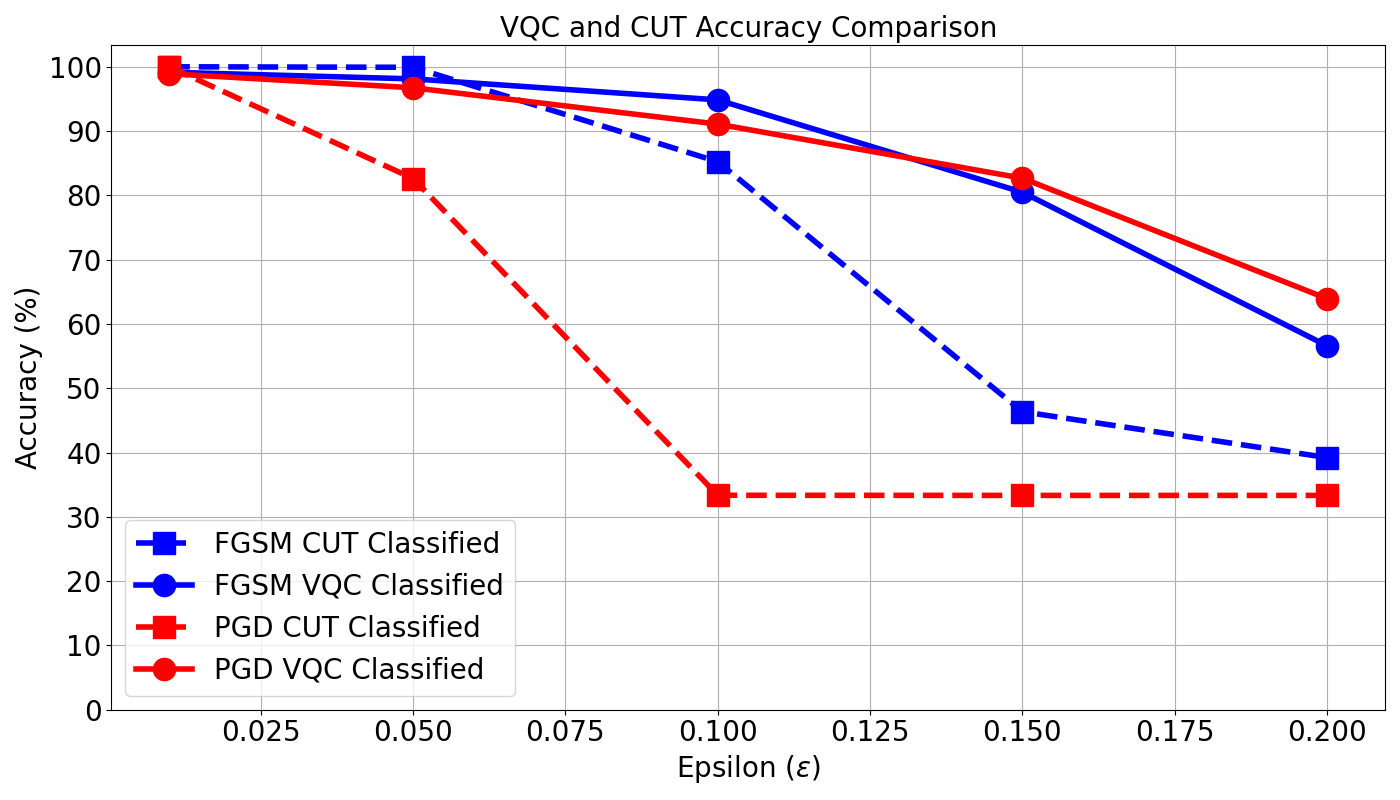}
\vspace{-2mm}
\caption{Accuracy vs the attack strength $\epsilon$ for VQC and CUT. 
}\vspace{-3mm}
\label{fig:vqc_cut}
\end{figure}

As mentioned, the VQ layer $\Omega$ in the VQC architecture is directly influenced by the CUT configuration, as the desired bottleneck size $d$ and the feature extractor's (FE) output dimension jointly determine the codeword length $\ell$. We here present the results for a given CUT, and for the combination of $N$ and $\ell$ that worked best for VQVAE, out of the pairs that we tested.
Under clean conditions, the CUT achieves a baseline accuracy of 100\%, slightly higher  compared to 99.47\% for the VQC. This minor difference can be attributed to the representational loss of quantization.
However, as the attack strength ($\epsilon$) increases, the benefits of VQ become evident. For very small $\epsilon$, CUT marginally outperforms VQC, but as $\epsilon$ grows, the VQC begins to dominate across all attack types. For instance, under a PGD attack, the VQC maintains approximately 90\% accuracy at $\epsilon = 0.1,$ while the CUT drops to around 33\%, effectively collapsing under stronger perturbations, as shown in Fig.\ref{fig:vqc_cut}. 

\begin{figure}[h]
\centering
\hspace{-1mm}\includegraphics[height=9cm, width=0.48\textwidth]{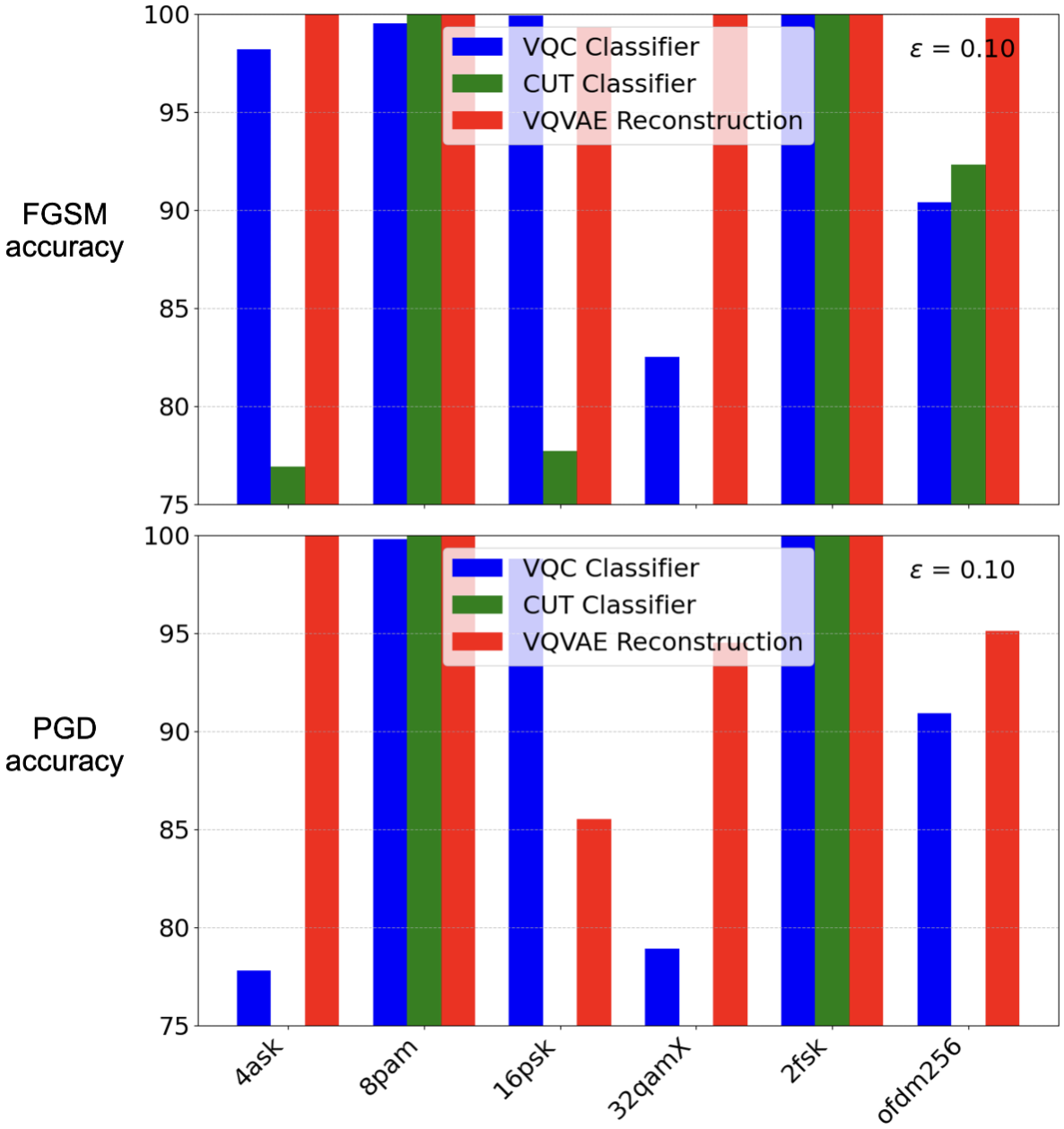}
\vspace{-2mm}
\caption{Comparing accuracy per class for $\epsilon = 0.1$: VQC, CUT, VQVAE; FGSM (top) vs. PGD.
}\vspace{-4mm}
\label{fig:vqc_matrix}
\end{figure}

This trend suggests that, while the VQ layer incurs a minor cost in clean-data performance, it provides substantial gains in adversarial robustness. The discrete bottleneck in the VQC acts as a noise filter, and reduces the model’s sensitivity to small, adversarially induced perturbations. In contrast, the continuous latent space of the CUT offers higher fidelity under gentle conditions but leaves the model more vulnerable to adversarial shifts. The robustness of the VQC model, in comparison to CUT and VQVAE-VIB, is further detailed for both FGSM and PGD through the class-specific accuracy graph at $\epsilon = 0.1,$  in Fig.\ref{fig:vqc_matrix}. Note that the $y$-axis starts with an accuracy of 75\%.  While overall accuracy remains above 90\% for both FGSM and PGD attacks, misclassifications occur primarily among specific modulation classes. VQVAE is more robust for 32QAM-X, and VQC for 16PSK. While our prior work \cite{AttaFCN} shows that  VQVAE preserves core signal features under perturbations, Fig.\ref{fig:vqc_matrix} and Fig.\ref{fig:set_dist} indicate more complex differences between VQ in VQVAE and VQC, and across modulation classes. Future work will provide evidence that VQVAE and VQC achieve optimal behavior in different discrete spaces (note that $N$ and $\ell$ were chosen to favor VQVAE). VQC may also require more training epochs for optimal robustness.

\subsection{VQVAE Performance}
\subsubsection*{\bf Quantization Effect: Comparing VQVAE and VAE}
\begin{figure}[h]
\vspace{-2mm}
\centering
\hspace{-1mm}\includegraphics[width=0.46\textwidth]{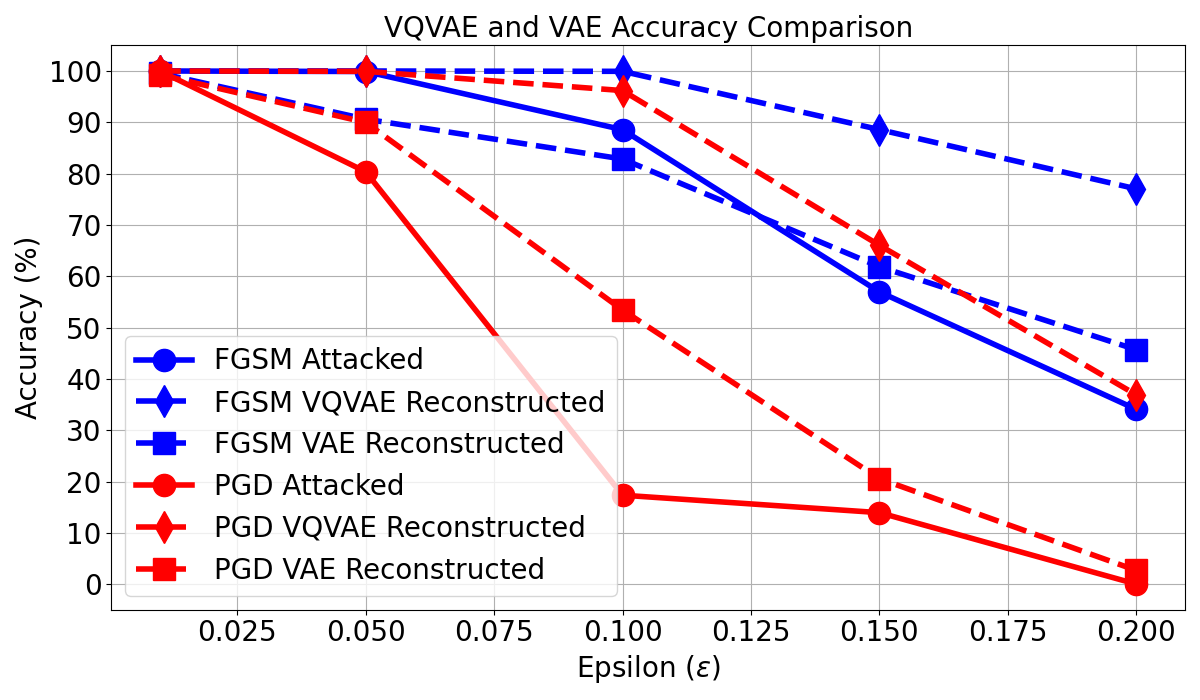}
\vspace{-2mm}
\caption{Accuracy vs the attack strength $\epsilon$ for VQVAE and VAE. 
}\vspace{-4mm}
\label{fig:vqvae_vae}
\end{figure}

We compare VAE and VQVAE to evaluate the effect of the discrete quantization bottleneck on adversarial robustness. On clean waveforms, both models reconstruct the input reasonably well, though the VAE may capture finer details due to its continuous latent space. Under adversarial attacks, differences emerge. The VQVAE consistently allows for high classification accuracy across multiple attack types, achieving over 95\% accuracy for FGSM and PGD attacks (and other tested attacks not presented here) for small to medium pertubations as shown in Fig.~\ref{fig:vqvae_vae}. In contrast, the VAE’s performance degrades more noticeably under attack: accuracy drops to 84\% for FGSM, and 54\% for PGD at $\epsilon =0.1$. These results highlight the advantage of DSVIB in the VQVAE, which filters out adversarial variations better than a VIB-trained VAE. 
\vspace{-1.5mm}
\section{Conclusion}
We investigated how to mitigate attacks on RF baseband waveforms using the VIB implemented in discrete latent spaces. We proposed two models: VQVAE, which filters adversarial perturbations before classification to restore the accuracy of the CUT, and VQC, which directly classifies signals through a discrete bottleneck. Both models were trained using a loss function directly derived from the general information bottleneck objective by including the task objective and stochastic vector quantization.

Experimental results demonstrate that DSVIB significantly improves robustness against tested white-box attacks compared to CUT and VAE of matching designs.  VQVAE achieves slightly higher robustness under the tested configurations, while VQC may require a different codebook size and longer training.  Future work will study these differences in more detail, with an increased number of samples and a wider variety of attacks. Latent space analysis requires additional research, as it indicates that DSVIB constrains perturbations differently in different modulated waveform types. Our findings confirm the potential of DSVIB for secure RF processing in Next Generation communications. 
\vspace{-1.5mm}
\bibliographystyle{IEEEtran}
\bibliography{advbib}

@INPROCEEDINGS{ScoreDiffChannel,
  author={Arvinte, Marius and Tamir, Jonathan I.},
  booktitle={IEEE Wireless Comm. and Network. Conference (WCNC)}, 
  title={{Score-Based Generative Models for Robust Channel Estimation}}, 
  year={2022},
}

@article{torchsig,
title={{Large Scale Radio Frequency Signal Classification}},
author={{L. Boegner et al.}},
year={2022},
journal=arXiv,
url={https://arxiv.org/abs/2207.09918}
}

@article{KingmaVAE,
author    = {Diederik P. Kingma and Max Welling},
title     = {An Introduction to Variational Autoencoders},
journal   = {Found. Trends Mach. Learn.},
volume    = {12},
number    = {4},
year      = {2019}
}

@article{vqvae2017neural,
  title={Neural discrete representation learning},
  author={{Van Den Oord, Aaron at al.}},
  journal={Advances in neural information processing systems},
  volume={30},
  year={2017}
}

@article{MIMOChannelEstimScore,
  author={Arvinte, Marius and Tamir, Jonathan I.},
  journal={IEEE Trans. on Wireless Comms}, 
  title={{MIMO Channel Estimation Using Score-Based Generative Models}}, 
  year={2023},
  volume={22},
  number={6},
  pages={3698-3713},
  }

@article{yu2024ai,
  title={{AI and Deep Learning for THz Ultra-Massive MIMO: From Model-Driven Approaches to Foundation Models}},
  author={{Yu, Wentao et al.}},
  journal={arXiv preprint arXiv:2412.09839},
  year={2024}
}

@ARTICLE{realizationRFadvers,
  author={{Kim, Brian et al.}},
  journal={IEEE Trans. on Wireless Communications}, 
  title={Channel-Aware Adversarial Attacks Against Deep Learning-Based Wireless Signal Classifiers}, 
  year={2022},
  volume={21},
  number={6},
  pages={3868-3880},
  }

@INPROCEEDINGS{Adexdetect,
  author={Kokalj-Filipovic, Silvija and Miller, Rob and Vanhoy, Garrett},
  booktitle={IEEE Global Conf. on Signal and Inform. Processing (GlobalSIP)}, 
  title={Adversarial Examples in RF Deep Learning: Detection and Physical Robustness}, 
  year={2019},
  volume={},
  number={},
  }

@INPROCEEDINGS{AEagainst,
  author={{Kokalj-Filipovic, Silvija et al.}},
  booktitle={Int. Conf. on Military Comms and Inf. Systems (ICMCIS)}, 
  title={Mitigation of Adversarial Examples in RF Deep Classifiers Utilizing AutoEncoder Pre-training}, 
  year={2019},
  volume={},
  number={},
  }

@article{sadeghi2018adversarial,
  title={Adversarial attacks on deep-learning based radio signal classification},
  author={Sadeghi, Meysam and Larsson, Erik G},
  journal={IEEE Wireless Communications Letters},
  volume={8},
  number={1},
  pages={213--216},
  year={2018}
}

@INPROCEEDINGS{dimred,
  author={Sahay, Rajeev and Mahfuz, Rehana and Gamal, Aly El},
  booktitle={Annual Conf. on Inform. Sciences and Systems (CISS)}, 
  title={Combatting Adversarial Attacks through Denoising and Dimensionality Reduction: A Cascaded Autoencoder Approach}, 
  year={2019},
}

@article{d2022dapps,
  title={dApps: Distributed applications for real-time inference and control in O-RAN},
  author={D'Oro, Salvatore and Polese, Michele and Bonati, Leonardo and Cheng, Hai and Melodia, Tommaso},
  journal={IEEE Communications Magazine},
  volume={60},
  number={11},
  pages={52--58},
  year={2022}
}

@misc{alemi2019deepvariationalinformationbottleneck,
      title={Deep Variational Information Bottleneck}, 
      author={Alexander A. Alemi and Ian Fischer and Joshua V. Dillon and Kevin Murphy},
      year={2019},
      eprint={1612.00410},
      archivePrefix={arXiv},
      primaryClass={cs.LG},
      url={https://arxiv.org/abs/1612.00410}, 
}

@article{singhLearnedComp,
  author    = {Saurabh Singh and
               Sami Abu{-}El{-}Haija and
               Nick Johnston and
               Johannes Ball{\'{e}} and
               Abhinav Shrivastava and
               George Toderici},
  title     = {End-to-end Learning of Compressible Features},
  journal   = {CoRR},
  volume    = {abs/2007.11797},
  year      = {2020},
  url       = {https://arxiv.org/abs/2007.11797},
}

@inproceedings{IBrobust,
  title={Improving adversarial robustness via information bottleneck distillation},
  author={Kuang, Huafeng and Liu, Hong and Wu, Yongjian and Satoh, Shin'ichi and Ji, Rongrong},
  booktitle={NeurIPS},
  year={2023}
}

@INPROCEEDINGS{AdvRobICMLCN1,
  author={{A. Furutanpey et al.}},
  booktitle={2025 IEEE Int. Conf. on Mach. Learning for Commun. and Networking (ICMLCN)}, 
  title={Adversarial Robustness of Bottleneck Injected Deep Neural Networks for Task-Oriented Communication}, 
  year={2025},
}

@inproceedings{ShannonCoding,
  title={Coding theorems for a discrete source with a fidelity criterion},
  author={C. E. Shannon},
  booktitle={IRE National Convention Record},
  volume={4},
  year={1959}
}

@inproceedings{TishbyIB,
author={N. Tishby and F. C. Pereira and W. Bialek}, 
title={{The information bottleneck method}},
booktitle={{37-th Annual Allerton Conf. on Commun., Control and Computing}}, 
year=1999}

@article{EstimI,
title={{Estimating mutual information}},
author={{A. Kraskov and H. Stögbauer and P. Grassberger}},
journal={Phys. Rev. E Stat. Phys., Plasmas, Fluids, and Related Interdisc. Topics},
volume={69},
number={6},
pages={16},
year={2004}
}

@inproceedings{AttaFCN,
author={A. Garuso and S. Kokalj-Filipovic and  Y. Kaasaragadda},
title={{A Look at Adversarial Attacks on Radio Waveforms from Discrete Latent Space}}, 
booktitle={{Conf. on Future Communication Networks, FCN}},
year=2025}

@InProceedings{OriginalModRec,
author={O'Shea, Timothy J.
and Corgan, Johnathan
and Clancy, T. Charles},
title={Convolutional Radio Modulation Recognition Networks},
booktitle={Engineering Applications of Neural Networks},
year=2016,
}

@article{ZHANG2022AMC,
title = {Deep learning based automatic modulation recognition: Models, datasets, and challenges},
journal = {Digital Signal Processing},
volume = {129},
pages = {103650},
year = {2022},
author = {Fuxin Zhang and Chunbo Luo and Jialang Xu and Yang Luo and Fu-Chun Zheng},
}

@ARTICLE{Access2021AMC,
  author={Huynh-The, Thien and Pham, Quoc-Viet and Nguyen, Toan-Van and Nguyen, Thanh Thi and Ruby, Rukhsana and Zeng, Ming and Kim, Dong-Seong},
  journal={IEEE Access}, 
  title={{Automatic Modulation Classification: A Deep Architecture Survey}}, 
  year={2021},
  volume={9},
  pages={142950-142971},
  }
\end{document}